\documentclass[journal]{IEEEtran}
\usepackage{cite}
\usepackage{amsmath,amssymb,amsfonts}
\usepackage{graphicx}
\usepackage{pgfplots}
\usepackage{url}

\usepackage{xcolor}
\usepackage{multirow}
\usepackage[nolist,printonlyused]{acronym}
\usepackage{verbatim}
\usepackage{url}
\usepackage{subcaption}
\usepackage{tabularx}
\usepackage{todonotes}
\usepackage{csquotes}

\usepackage{tikz}
\usetikzlibrary{shapes,arrows,calc,positioning,decorations.pathmorphing} 
\tikzset{
  -|-/.style={
    to path={
      (\tikztostart) -| ($(\tikztostart)!#1!(\tikztotarget)$) |- (\tikztotarget)
      \tikztonodes
    }
  },
  -|-/.default=0.5,
  |-|/.style={
    to path={\textbf{}
      (\tikztostart) |- ($(\tikztostart)!#1!(\tikztotarget)$) -| (\tikztotarget)
      \tikztonodes
    }
  },
  |-|/.default=0.5,
}
\tikzset{
   block/.style = {draw, rectangle,
       minimum height=1cm,
       align = center
   },
   input/.style = {coordinate,node distance=1cm},
   output/.style = {coordinate,node distance=1cm},
   arrow/.style={draw, -latex,node distance=2cm},
   pinstyle/.style = {pin edge={latex-, black,node distance=2cm}},
   sum/.style = {draw, circle, node distance=1cm},
   gain/.style = {
     regular polygon, regular polygon sides=3,
     draw, fill=white, text width=1em,
     inner sep=0mm, outer sep=0mm,
     shape border rotate=-90
   },
   dot/.style={circle,fill,draw,inner sep=0pt,minimum size=3pt}
 }

\usepackage{pgfplots}
\pgfplotsset{compat=1.9}

\newenvironment{myaxis}[1][2]{%
\begin{axis}[%
    xlabel={Time (s)},
    x label style={at={(axis description cs:0.5,-0.11)},anchor=north},
    ylabel={Avg. reputation},
    y label style={at={(axis description cs:-0.22,0.5)},anchor=north},
    ymajorgrids,
    xmajorgrids,
    ymax = 2600, 
    grid style=dashed,
    width=\textwidth,
    height=5cm,
    xtick distance=900,
    no markers,
    legend columns=#1 
]
}{
\end{axis}
}

\newenvironment{myaxis1}[1][2]{%
\begin{axis}[%
    xlabel={Time (s)},
    x label style={at={(axis description cs:0.5,-0.11)},anchor=north},
    ylabel={Avg. reputation},
    y label style={at={(axis description cs:-0.22,0.5)},anchor=north},
    ymajorgrids,
    xmajorgrids,
    ymax = 3400, 
    grid style=dashed,
    width=\textwidth,
    height=4.5cm,
    xtick distance=900,
    no markers,
    legend columns=#1 
]
}{
\end{axis}
}

\newcommand{\myparamstudyplot}[2]{%
\addplot table [x=time, y=TotalReputationAverage] {#1};
\addlegendentry{#2}
}

\newcommand{\myattackplot}[1]{%
\addplot table [x=time, y=ActiveType1ReputationAverage] {#1};
\addlegendentry{Legitimate}
\addplot[orange] table [x=time, y=ActiveType2ReputationAverage] {#1};
\addlegendentry{Malicious}
}

\newcommand{\mycoordinatedattackplot}[1]{%
\addplot table [x=time, y=ActiveType1ReputationAverage] {#1};
\addlegendentry{Legitimate}
\addplot table [x=time, y=ActiveType2ReputationAverage] {#1};
\addlegendentry{Malicious}
}

\newcommand{\myparagraph}[1]{\noindent\textbf{#1}}

\usepackage{xspace}
\newcommand{\mypar}[1]{\smallskip\noindent\textbf{#1.}\xspace}

\newcommand{\systemname}{\emph{GOLIATH}\xspace}

\title{\systemname: A Decentralized Framework for Data Collection in Intelligent Transportation Systems}

\author{Davide~Maffiola, 
        Stefano~Longari, 
        Michele~Carminati, 
        Mara~Tanelli,~\IEEEmembership{Senior Member,~IEEE,}
        and~Stefano~Zanero,~\IEEEmembership{Senior Member,~IEEE.}
\thanks{ All authors are with the Dipartimento di Elettronica, Informazione e Bioingegneria, Politecnico di Milano, Italy. E-mail: davide.maffiola@mail.polimi.it \{stefano.longari, michele.carminati, mara.tanelli, stefano.zanero\}@polimi.it}}

\begin{acronym}
	{\small
	\acro{CPS}{Cyber-Physical System}
	\acro{TEE}{Trusted Execution Environment}
	\acro{CAN}{Controller Area Network}
	\acro{CAN-FD}{Controller Area Network with Flexible Data-rate}
	\acro{DSRC}{Dedicated Short Range Communications}
	\acro{ECU}{Electronic Control Unit}
	\acro{TCU}{Telematic Control Unit}
	\acro{OBD-II}{On-Board Diagnostic}
	\acro{OBD}{On-Board Diagnostic}
	\acro{IoT}{Internet of things}
	\acro{OEM}{Original Equipment Manufacturer}
	\acro{MOST}{Media Oriented System Transport}
	\acro{ADAS}{Advanced Driver Assistance System}
	\acro{ABS}{Antilock Braking System}
	\acro{LIN}{Local Interconnect Network}
	\acro{TPMS}{Tire Pressure Monitoring System}
	\acro{UDS}{Unified Diagnostic System}
	\acro{IDS}{Intrusion Detection System}
	\acro{ADS}{Anomaly Detection System}
	\acro{LSTM}{Long Short-Term Memory}
	\acro{OTA}{Over The Air}
	\acro{RNN}{Recurrent Neural Network}
	\acro{HMM}{Hidden Markov Model}
	\acro{ACK}{Acknowledgment}
	\acro{CRC}{Cyclic Redundancy Check}
	\acro{IFS}{Interframe Space}
	\acro{TEC}{Transmit Error Count}
	\acro{REC}{Receive Error Count}
	\acro{LTE}{Long Term Evolution}
	\acro{DoS}{Denial of Service}
	\acro{BC}{Bit Counter}
	\acro{PC}{Polarity Counter}
	\acro{SoF}{Start of Frame}
	\acro{DLC}{Data-Length Code}
	\acro{IDE}{Identifier Extension}
	\acro{RPM}{Repetiition Per Minute}
	\acro{RTR}{Remote Transmission Request}
	\acro{EoF}{End of Frame}
	\acro{IF}{Intermission Field}
	\acro{DL}{Deep Learning}
	\acro{ML}{Machine Learning}
	\acro{OCSVM}{One-Class SVM}
	\acro{CAN-H}{CAN High}
	\acro{CAN-L}{CAN Low}
	\acro{CRF}{Conditional Random Field}
	\acro{ROC}{Receiver Operating Characteristic Curve}
	\acro{AUC}{Area Under the Curve}
	\acro{rpm}{repetition per minutes}
	\acro{GRU}{Gated Recurrent Unit}
	\acro{PID}{Proportional, Integral and Derivative}
	
	\acro{ITS}{Intelligent Transportation System}
	\acro{IVI}{In-Vehicle Infotainment}
	\acro{V2X}{Vehicle to Everything}
	\acro{C-V2X}{Cellular V2X}
	\acro{RSU}{Road Side Unit}
	\acro{FCD}{Floating Car Data}
	\acro{GPS}{Global Positioning System}
	\acro{GNSS}{Global Navigation Satellite System}
	\acro{P2P}{Peer-to-Peer}
	\acro{PoW}{Proof-of-Work}
	\acro{PoS}{Proof-of-Stake}
	\acro{PBFT}{Practical Byzantine Fault Tolerance}
	\acro{FCC}{Federal Communications Commission}
	\acro{PRNG}{Pseudorandom Number Generator}
	\acro{MoST}{Monaco SUMO Traffic}
	\acro{GSM}{Global System for Mobile Communications}
	\acro{PoI}{Proof-of-Importance}
	\acro{PoB}{Proof-of-Believability}
	\acro{SDVN}{Software Defined Vehicular Network}
	}
\end{acronym}

\begin{document}
\maketitle

%

\begin{abstract} 
\acp{ITS} technology has advanced during the past years, and it is now used for several applications that require vehicles to exchange real-time data, such as in traffic information management. Traditionally, road traffic information has been collected using on-site sensors. However, crowd-sourcing traffic information from onboard sensors or smartphones has become a viable alternative. State-of-the-art solutions currently follow a centralized model where only the service provider has complete access to the collected traffic data and represent a single point of failure and trust.

In this paper, we propose \systemname, a blockchain-based decentralized framework that runs on the \ac{IVI} system to collect real-time information exchanged between the network's participants. 
Our approach mitigates the limitations of existing crowd-sourcing centralized solutions by guaranteeing trusted information collection and exchange, fully exploiting the intrinsic distributed nature of vehicles. 

We demonstrate its feasibility in the context of vehicle positioning and traffic information management. Each vehicle participating in the decentralized network shares its position and neighbors' ones in the form of a transaction recorded on the ledger, which uses a novel consensus mechanism to validate it. We design the consensus mechanism resilient against a realistic set of adversaries that aim to tamper or disable the communication. 
We evaluate the proposed framework in a simulated (but realistic) environment, which considers different threats and allows showing its robustness and safety properties.   
\end{abstract}



\section{Introduction}

\noindent In the last years, \acfp{ITS} have gone through significant changes, mainly due to the introduction of new technologies deployed on vehicles, which allow advanced applications that often require vehicles to exchange real-time data to provide high-quality services. 
Modern vehicles are now equipped with \acf{IVI} systems that allow executing custom applications on top-notch mobile processors~\cite{audi-exynos}. 
Additionally, vehicle manufacturers are investing on \ac{V2X} communication~\cite{abboud2016interworking}, either based on cellular network~\cite{audi-5g} or on short range communication technologies~\cite{volkswagen-dsrc}.
 These features allow vehicles to perform complex operations and take part in a vehicular network, thus enabling the deployment of lightweight blockchain-based solutions, which guarantee trust without relying on a single, centralized point of failure.

The most common field of application of \acp{ITS} is traffic information management, which requires the processing of a high volume of real-time traffic information. Currently, traffic information sources are local authorities or agencies that use costly on-site sensors such as \acp{RSU}, or services based on \ac{FCD} and crowdsourcing like \emph{Waze}, \emph{TomTom HD Traffic}, and \emph{Google Maps}. However, both technologies are based on centralized systems, where the service providers have complete control of the collected data and acts as a trusted element. Although these methods can provide crowdsourced traffic information, they suffer from high deployment and maintenance costs, and their effectiveness is limited to a specific area. Also, precise traffic information is challenging to measure and estimate, and the available sources are often partial or not easily accessible. In fact, in last years we assisted to the growth of blockchain-based solutions applied to various domains of the automotive ecosystem~\cite{yahiatene2018towards,yahiatene2019blockchain,yang2017blockchain,yang2018blockchain,li2018creditcoin,jiang2018blockchain,distefano2021trustworthiness,javaid2019drivman,sharma2018blockchain,li2019toward,huang2018lnsc,lei2017blockchain,wang2020secure,kang2017enabling,liu2018blockchain,hassija2020blockchain,knirsch2018privacy}. However, these solutions are usually only semi-decentralized since they rely on base stations (e.g., \acp{RSU}) and are not evaluated against advanced attacks in the automotive environment.  

In this paper, we aim to solve the challenges mentioned above by proposing \systemname, a blockchain-based framework that relies on the resources offered by \ac{IVI} systems to provide a decentralized alternative to centralized services commonly used to collect real-time information exchanged between network's participants. We demonstrate its feasibility in the context of vehicle positioning and traffic information management by processing real-time traffic information. 
In the proposed framework, each participating vehicle shares its position and its neighbors' ones, obtained through an on-board \ac{GNSS} module, in the form of blockchain transactions. To detect nearby vehicles, the participants periodically use short-range communication to broadcast a probe message. When the participants receive a reply to their probe message, they generate, sign, and broadcast, for each received answer, a transaction containing the information about the occurred interaction (i.e., the probing entity's position and the vehicles' identifiers). 
The transactions are then collected in a block, which goes through a novel consensus mechanism for validation. The consensus mechanism verifies the integrity of the received information and evaluates the participants' reliability and trustworthiness. 
By doing so, \systemname is resilient against adversaries aiming to tamper or disable the communication, which instead have been proven effective against current centralized solutions~\cite{weckert2020google}. Once validated,  all participants are updated with the new block. The consensus mechanism is entirely managed by the participant vehicles with supporting \acp{RSU} and does require any entity to have special privileges or permissions after the break-in phase. 
This allows the framework to be fully decentralized, implying that a trusted intermediary -- i.e., the service provider -- is no longer required. 

We validate \systemname by implementing a network simulator for realistic traffic generation. We demonstrate the resilience of the framework in realistic scenarios while also showing its capability of maintaining high-quality traffic information even in the presence of malicious entities.

Our contributions are the following: 
\begin{itemize}
\item \systemname, a lightweight, decentralized, blockchain-based framework for real-time data collection in \acp{ITS} that mitigates the limitation of centralized systems and fully exploits the intrinsic distributed nature of vehicles. We also demonstrate its feasibility in the traffic information management domain. 
\item A novel consensus mechanism designed to be resilient, differently from existing centralized solutions~\cite{weckert2020google}, against a realistic set of adversaries that aim at tampering or disabling the communication. The consensus mechanism is entirely managed by the participant vehicles, removing the need for a centralized service provider.
\item A simulation tool (available at \url{https://github.com/necst/GOLIATH}) which is, to the best of our knowledge, a first attempt to evaluate decentralized traffic information management systems through the simulation of realistic attack scenarios. This tool will help future researchers in the comparison with the state of the art and in the evaluation of the robustness of their approaches.
\end{itemize}

The paper is structured as follows: In Section~\ref{sec:problem} we describe the problem we aim to tackle and our research goal. In Section~\ref{sec:background} we offer an overview of blockchain technologies, while in Section~\ref{sec:related}, we discuss the current status of positioning systems and related works that study the blockchain technologies in the automotive domain. Then, in Section~\ref{sec:threat} we provide an analysis of the threat model, fundamental to understand the choices we made in Section~\ref{sec:approach}, where we present \systemname. In Section~\ref{sec:experiments} we show the experimental evaluation of our framework. Finally, in Section~\ref{sec:conclusion} we discuss the conclusions of our work, discussing the main limitations and outlining some future works.


\section{Background on Blockchain Technology}
\label{sec:background}

\noindent Blockchain is a vast research topic, for which we refer the reader to \cite{blockchainsurveymonrat, zheng2017overview}. In this section, we provide the basic concepts needed to understand our contribution. 

A blockchain is a data structure employed in decentralized and distributed systems to obtain immutable and verifiable data in the form of blocks. This solves the issues deriving from the lack of a centralized trusted intermediary intrinsic of a \ac{P2P} network. A blockchain-based system is composed of three core components: \textbf{nodes}, which are the participants of the network, \textbf{transactions}, which represents the interactions between nodes (e.g., monetary transactions in blockchain-based cryptocurrencies), and \textbf{blocks}, which are collections of valid transactions over a period of time. 
On top of these three components, blockchain-based systems require the definition of two processes to ensure authenticity and verification of data: the first is a \textbf{signature scheme} that grants the integrity and authenticity of transactions and blocks, and the second is a \textbf{consensus mechanism}, which has the goal of validating data. The consensus mechanism is composed of a \textbf{consensus algorithm} and a \textbf{validation algorithm}. The former is used to achieve a distributed consensus, and the latter checks the validity of the content of a block. 
One of the most known consensus algorithms is the \ac{PoW}, implemented initially by Bitcoin~\cite{nakamoto2008bitcoin}, which chooses the next valid block by rewarding nodes based on expended computation and time. A second class of consensus algorithms is the \ac{PoS}~\cite{king2012ppcoin}, where instead of computational power, an abstract resource imaginable as a currency is used by the nodes as a stake. The higher the stake a node offers, the higher is the chance of that node generating the next block. In this work, we opt for a \ac{PoS}-based system in light of the strict computational requirements of the automotive field that would make \ac{PoW} unfeasible. 

From a high-level perspective, nodes (i.e., participants) of a blockchain generate transactions that express the interactions between each other and broadcast them. 
Harvesters then collect these transactions into candidate blocks and compete for the right of adding their candidate block to the blockchain through the consensus algorithm designed for that specific blockchain system.
The block of the winning harvester is then added to the blockchain by attaching to it the hash of the previous block, officially chaining the new block to the blockchain.
The robustness of a blockchain-based system depends on the number of legitimate participants in the network since they directly influence the consensus mechanism. Hence, it is of paramount importance for a blockchain-based system to reach critical mass (i.e., a high number of participants) in a short period and that the consensus mechanism rewards legitimate behaviors only.

\section{Related Works}
\label{sec:related}

\noindent In the following section, we present related blockchain-based works that pave the way for the validity of our approach and, then, we illustrate decentralized solutions applied to the automotive field.

\subsection{Related Blockchain Methodologies}
\noindent  In the last ten years, the need for decentralized solutions to share information has been answered most of the time by the blockchain technology. 
Initially in 2008 by Nakamoto with Bitcoin\cite{nakamoto2008bitcoin}, followed by many others. We briefly discuss those that apply techniques related to or similar to ours. ByzCoin~\cite{kogias2016enhancing} proposes to improve the transaction acceptance delay of Bitcoin through the use of a consensus mechanism based on \ac{PBFT}~\cite{castro1999practical}, where view-based leader election is decoupled from the validation algorithm. The \ac{PoB} consensus algorithm used in IOST~\cite{iost2017iost}, which is a modification of a \ac{PoS} algorithm, divides the participants into a believable league and a normal league, similarly to the method used in the proposed framework. The believable league, though, is only used to validate the transactions optimistically, and the normal league still has to run the modified version of \ac{PBFT} used in ByzCoin. 
In particular, the blockchain technology has been applied in nearly all domains of the IoT~\cite{yahiatene2019blockchain}, from the healthcare~\cite{liang2017integrating,esposito2018blockchain,guo2018secure} and embedded devices~\cite{lee2017blockchain,gu2018consortium}, to fog computing~\cite{huang2018bitcoin} and software-defined networking~\cite{sharma2017distblocknet}. 

\subsection{Related Blockchain-based works in the automotive field}
\noindent Blockchain-based solutions have also been applied to the automotive ecosystem. In particular, the most common applications are related to trust-based dynamic data collection and exchange~\cite{yahiatene2018towards,yahiatene2019blockchain,yang2017blockchain,yang2018blockchain,li2018creditcoin}, secure storage of static data~\cite{jiang2018blockchain,distefano2021trustworthiness,javaid2019drivman,sharma2018blockchain,li2019toward}, key management for authentication and verification~\cite{huang2018lnsc,lei2017blockchain,wang2020secure}, and energy trading in vehicular networks~\cite{kang2017enabling,liu2018blockchain,hassija2020blockchain,knirsch2018privacy}.

Regarding the trust-based dynamic data collection and exchange domain, to which our work belongs, the blockchain is used for exchanging information in dynamic vehicular networks (i.e., traffic management systems). The security (e.g., integrity and availability) of the exchanged information is guaranteed through the concept of trust among the entities involved in the communication, where only entities that gain \textit{trust} through legitimate interactions and can participate in the blockchain. 
Yahiatene et al.~\cite{yahiatene2018towards,yahiatene2019blockchain} propose a framework based on \ac{SDVN} and blockchain that builds trust through the feedback of satisfaction of other nodes regarding the received information. The process is handled in a semi-centralized manner by \acp{RSU}, and the miners' election is done dynamically through a novel distributed miners-connected dominating set algorithm (DM-CDS).
Yang et al.~\cite{yang2017blockchain,yang2018blockchain} propose a trust model based on blockchain where the information received by the \ac{RSU}, which acts as a centralized controller, is given a credibility rating that builds the trust value of each of the nodes. 
Li et al.~\cite{li2018creditcoin} propose Creditcoin, a framework to encourage privacy-preserving communication of alerts between vehicles with a cryptocurrency-like solution, where \acp{RSU} are used as a trusted element, and the alert generator is rewarded by receiving a tailored currency in a \ac{PoS}-like fashion.

Regarding the secure storage of static data domain, multiple solutions have been designed to securely and publicly share long-term data from different sources, such as vehicle and parts chain of ownership, insurance data, and other VANET-related data (e.g., driving habits, on-board sensors data), through the use of various blockchain technologies. Distefano et al.~\cite{distefano2021trustworthiness}, Jiang et al.~\cite{jiang2018blockchain}, Sharma et al.~\cite{sharma2018blockchain}, and Javaid et al.~\cite{javaid2019drivman} study the feasibility and the requirements needed to build blockchain-based frameworks to accommodate the needs of the automotive domain, while Li et al.~\cite{li2019toward} tackle the specific issue of ad dissemination in vehicular networks.

The works that fall in the key management for authentication and verification domain~\cite{huang2018lnsc,lei2017blockchain,wang2020secure}, as the name suggests, tackle the issue of authentication in vehicular networks, which have tailored requirements such as privacy-preserving pseudonyms, fast key transfer time, and low computing and communication overheads. 

Finally, multiple works focus on the issue of energy trading in vehicular networks~\cite{kang2017enabling,liu2018blockchain,hassija2020blockchain,knirsch2018privacy}, handling the exchange of data required to track the involved entities and transactions, again focusing on the preservation of the privacy of the vehicle owners while considering the requirements of vehicular networks.

\mypar{Discussion on related works} In this paper, we propose a blockchain-based framework for real-time data collection in \acp{ITS} that exploits a novel consensus mechanism that falls into the trust-based dynamic data collection and exchange domain. It is evident how the scope and structure of solutions that fall in the other aforementioned categories, although being based on blockchain technologies, significantly differ from the approach of our paper, making a direct comparison unfeasible. 
Regarding the trust-based dynamic data collection and exchange domain, all the presented related works manage participants (e.g., miners, harvesters) differently. In fact, most of the time, they are based on \acp{RSU} or base station to which the \textit{mining} process is delegated. Consequently, related works are based on a centralized or at most semi-decentralized architecture, which significantly differs from our approach, which is fully decentralized (the \acp{RSU} are needed only in the initialization phase). Indeed, the comparison of output results can not take place, besides being of little significance. Therefore, in line with the presented related works, we resort to demonstrating the robustness of the blockchain-based solution by studying the resistance against known attacks (e.g., Sybil attacks). In addition, in the presented work, we further evaluate the feasibility of our approach in the automotive scenario by exploiting a network simulator for realistic traffic generation, demonstrating the resilience of the framework in realistic scenarios while also showing its capability of maintaining high-quality traffic in-formation even in the presence of malicious entities. This contribution, in particular, is missing or only briefly discussed in all the above-presented works. 

\section{Problem Statement and Research Question}
\label{sec:problem}

\noindent Exchange of data between vehicles in \ac{V2X} environments has currently been designed mainly as a centralized, proprietary communication that uses technologies such as 4g, 5g, and \ac{DSRC} to transfer data. Regarding the context of vehicle positioning and traffic information, local authorities and agencies in large cities, e.g., Transport for London~\cite{tlf}, often provide real-time traffic information on an open-access basis. This information is primarily collected through on-site sensors like CCTV cameras, Wi-Fi data, and induction loops, possibly paired with artificial intelligence techniques~\cite{buch2009vehicle}. Several companies, like TomTom~\cite{rehborn2012traffic}, Google~\cite{barth2009bright}, and Waze~\cite{amin2018evaluating} have adopted crowdsourcing and \ac{FCD}~\cite{leduc2008road} to remove the need of on-site sensors. This approach relies on \ac{GNSS} and \ac{GSM} to collect the real-time positions from the devices of active users. 

Although these companies allow users to obtain aggregated traffic information, the raw data are not publicly accessible, and the offered services adopt a loose security model. These centralized systems have been proven vulnerable to several successful attacks that tamper with the collected traffic information, the most recent being from Simon Weckert~\cite{weckert2020google}.

\mypar{Research Question} Based upon the same technologies premises (i.e., crowdsourced information collection and global positioning), the research presented in this paper aims to explore the feasibility of a fully decentralized framework for vehicle data collection based on a blockchain data structure, whose properties directly tackle the weaknesses of current centralized solutions while preserving their functioning. To achieve the research goal, we have to face two challenges. 
The first challenge regards ensuring the system's resilience (i.e., reliable and trusted information collected) against realistic attacks.
The second challenge regards the compatibility of the proposed framework with existing automotive technology. In fact, computational power is a known constraint in the automotive domain.
Even if \systemname can be applied to any exchanged data -- with proper modification -- in the remainder of this work, we apply our approach to the traffic information management domain since it is one of the most common and best fits the problem under analysis. 
\section{Threat Model}
\label{sec:threat}

\noindent Threat modeling~\cite{shostack} is essential to derive the non-functional security requirements in the design of a system.

\myparagraph{Threat Agents.} The main threat agents are individuals (or small groups) who aim at profiting from disrupting the service (e.g., ransomware attack) or wish to hijack traffic for personal gain or to affect society at large. Attackers may also wish to forge information for anti-forensic purposes (i.e., claiming they are or they have been in a different position). Another potential threat comes from competitors who may gain an advantage from a loss of reputation of the targeted service. Therefore, the most significant threats are a disruption or degradation of service or targeted manipulation of the results.

\myparagraph{Attacker Goals.} 
To disrupt the service, an attacker could aim to lower the overall quality of the collected information (e.g., broadcasting a high number of crafted artifacts) or deploy a \ac{DoS} attack by exhausting the resources -- computational or network related -- of the participants (e.g., flooding the system with forged data).
To manipulate the system results or to forge fake records, attackers could aim to make the system accept as valid maliciously crafted data (e.g., faking the position of vehicles).  
It should be noted that the two goals affect each other since the robustness of blockchain-based solutions depends on the number of participants. Disrupting the service lowers the (perceived) reliability of the system, which in turn lowers the interest in participating of legitimate nodes. If a minor number of nodes participates in the consensus mechanism, the capability of attackers to manipulate the system increases. 

\myparagraph{Attack Modeling.} Attackers can perform their attacks on the blockchain core components: at a transaction level, block, or consensus level. 
At a transaction level, the attacker can forge or spoof fake transactions. To name a few, they can selectively reply only to \ac{DSRC} probes sent by some participants, or broadcast transactions with incorrect information, or send spam transactions. 
At a block level, attackers attempt to tamper with the blockchain validation algorithm by altering the validity tags of transactions, the reputation scores, or the activity flags. 
Also, attackers can try to gain the majority during the execution of the consensus algorithm and reject a legit block or force a forged block to be accepted. These block-level attacks fall into the category of Sybil attacks~\cite{douceur2002sybil}.

\section{ \systemname Approach} 
\label{sec:approach}

\begin{figure}
\centering
\includegraphics[width=0.9\columnwidth]{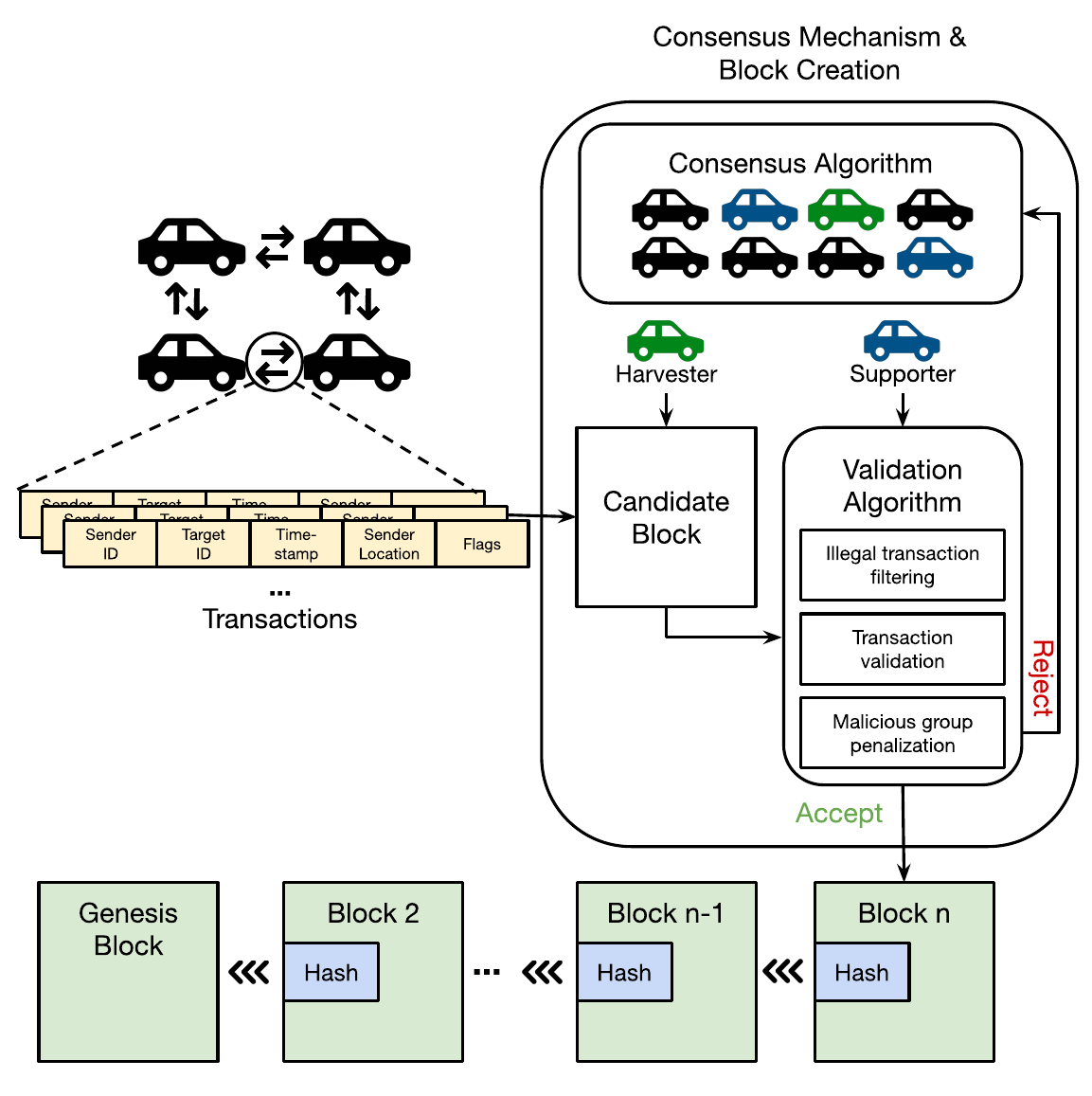}
\caption{Overview of \systemname Approach.}
\label{fig:approach-components}
\vspace{-10 pt}
\end{figure}

\noindent In Figure~\ref{fig:approach-components} we provide an overview of \systemname. 
Each participating vehicle shares (i.e., broadcasting a message) its position and its neighbors' ones in the form of blockchain transactions. 
In the meantime, following the consensus mechanism, the participants elect a harvester, which generates a valid candidate block by aggregating the broadcasted transactions, and a set of supporters, which run the validation algorithm. 
The candidate block, generated by the harvester, is evaluated by the supporters following the validation algorithm and, if valid, is broadcasted and added to the blockchain.  

The consensus mechanism determines how the blockchain grows by validating blocks. It performs primarily two operations. First, it executes a \emph{consensus algorithm} to ensure that the choice between accepting or rejecting a new block is indisputable by electing harvesters and supporters.  This operation permits decentralization and the removal of trusted intermediaries, allowing the consensus mechanism to work in a trust-less environment. 

The validation algorithm ensures the integrity of the received information and evaluates the participants' reputations. We define the \textit{reputation} as a measure of the reliability and trustworthiness of a participant that depends on the correctness of the history of its past transactions, and we use it as the abstract resource (i.e., stake) in our \ac{PoS}. In fact, participants may broadcast incorrect or malicious transactions. Therefore, validation is required to decide whether a transaction is valid or not, and the reputation of the transmitting node is updated accordingly. Unlike the consensus algorithm, the validation is domain-specific.

\subsection{Blockchain Data Structure}

\noindent The blockchain data structure stores traffic information in the form of transactions and blocks. It guarantees the properties of immutability, transparency, decentralization, persistence, and pseudo-anonymity~\cite{puthal2018everything}. 

\myparagraph{Transaction.} The primary entity in a blockchain data structure is the \emph{transaction}, representing the interaction between two or more participants. In our context, transactions are generated by the participants through short-range communication.
Participants -- also called nodes -- periodically broadcast a probe message that is captured by nearby vehicles (i.e., vehicles in communication range), which respond with a reply message. Regarding the communication range, it can be defined by both distance-bounding protocols~\cite{brands1993distance} or by considering the maximum detection distance of short-range communication devices. Since, according to the US \ac{FCC} classification, a low-power \ac{DSRC} transmitter has a maximum range of about 15 meters~\cite{kenney2011dedicated}, from this point on, we will consider as \enquote{short-range} a distance in the order of a few tens of meters. 
Then, the probing vehicles generate a \emph{transaction} for each reply received, which contains their identifier, the identifier of the detected participant, their position obtained through \ac{GNSS} (e.g., \acs{GPS}), and a timestamp. 
The sender digitally signs its transactions individually, making them immutable, and broadcasts them to all the other participants using a long-range communication mean (e.g., cellular network).

\myparagraph{Block.} The second element that composes the blockchain is the \emph{block}, which contains all the information that defines the current state of the blockchain and references the previous one through its hash value, forming a structure similar to a linked list. 
Each block contains up to a maximum number of transactions, which is defined by a \enquote{block-size}. 
Blocks are added to the blockchain after being validated by the consensus mechanism (see Section~\ref{sec:consensus}), and each participant is rewarded by increasing their reputation.

\begin{figure}
\centering
\includegraphics[width=0.9\linewidth]{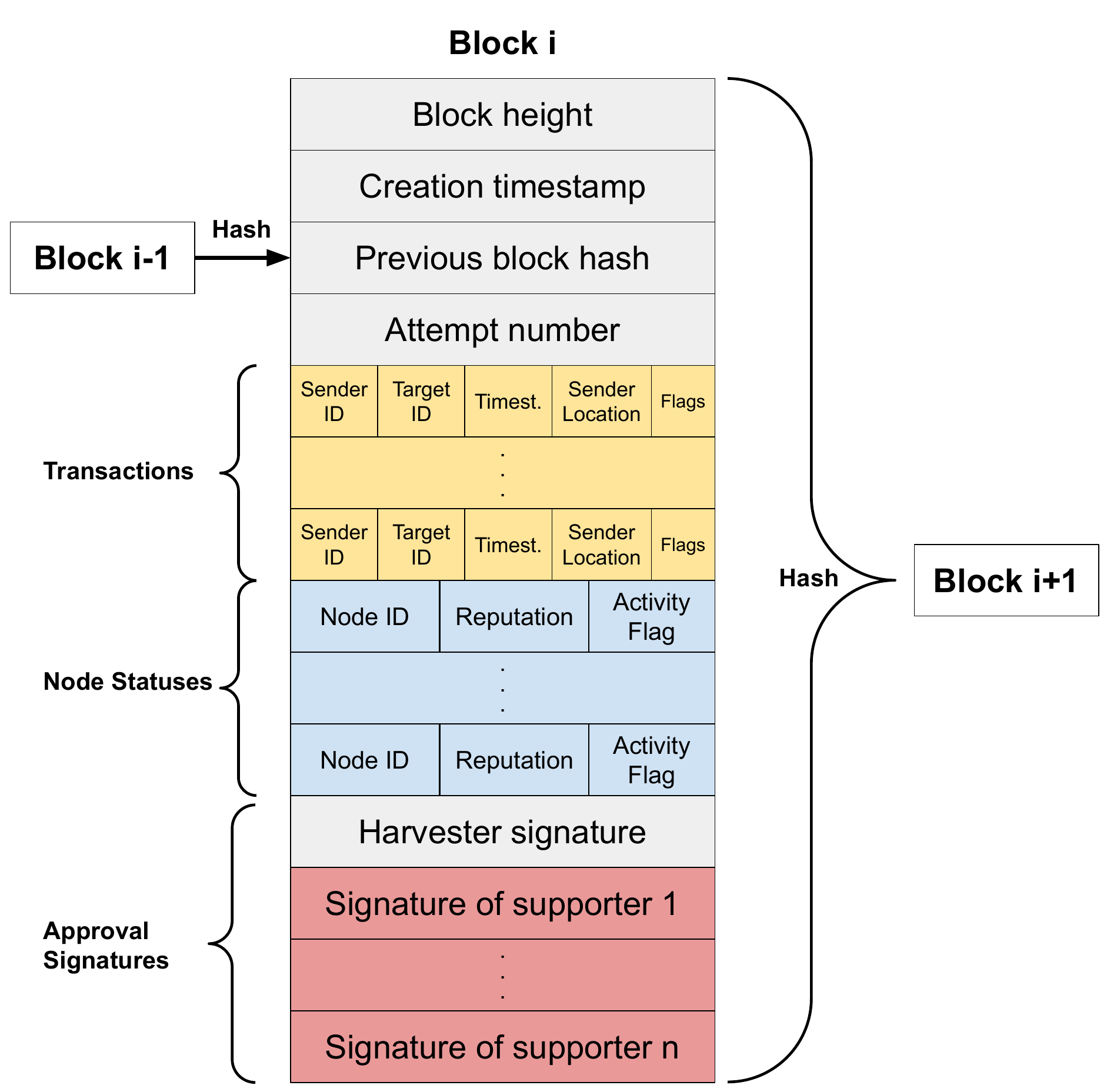}
\caption{Structure of the blockchain and of the blocks used in the proposed framework.}
\label{fig:block-content}
\vspace{-10 pt}
\end{figure}

A new block is inserted after a fixed amount of time from the previous insertion, which we refer to as \enquote{block-time} (i.e., the interval between two executions of the consensus mechanism). If more transactions than the block size are generated in the block-time, we apply a \enquote{greedy} selection strategy. When a new transaction is received, our strategy randomly selects a transaction from the participant with the highest number until it finds one that is older and removes it.  
Blocks also include a tag that indicates each transaction validity status, the list of identifiers of the participants, an integer score representing their reputation, and a flag that indicates if the participant has been active during the \enquote{block-time} (A node is considered active if at least one of its transactions is evaluated as valid).  
The metadata stored in the block includes a timestamp indicating when the block was created, the digital signature of the node that harvested the block, the block height, which is the index of the block from the genesis block (i.e., the first validated block), and the attempt number, that is derived from the elapsed time since the last block as $$a=\lfloor\frac{\Delta time}{block-time}\rfloor$$ 
The attempt number is necessary to distinguish different attempts when the consensus mechanism fails, causing a block to be rejected.
Figure~\ref{fig:block-content} graphically summarizes the structure of the blocks and the transactions.

\myparagraph{Genesis Block.} A crucial aspect to consider is the generation of the first block in the blockchain, which is called the \emph{genesis block}. It has to satisfy the assumptions of the validation phase (see Section~\ref{sec:validation}) and allows \systemname to start correctly. Consequently, the genesis block contains a set of randomly generated transactions between initially trusted participants (e.g., \acp{RSU}). In other words, the genesis block must be trusted in the break-in phase. After this phase, these trusted participants are no longer required and are treated as normal ones. 

\myparagraph{Resources Optimization.} The amount of memory required to store the whole blockchain grows constantly. This represents a critical issue for embedded systems, such as \acf{IVI} systems, where the amount of memory is limited. It is possible to introduce the concept of \emph{state blocks}~\cite{kokoris2018omniledger, iost2017iost}, leaving the task of memorizing the full blockchain only to the participants with more resources, like \acp{RSU}. When there are enough regular blocks, the consensus mechanism produces a new state block that summarizes and aggregates the oldest regular blocks, allowing participants with limited resources to store only state blocks and the most recent regular blocks. Moreover, new nodes that want to join the blockchain have to download and verify less data, granting a faster join.
 
\myparagraph{Key distribution and Pseudonymity.} The node identifiers and their respective asymmetric keys are bound to one vehicle. Their distribution may be performed by car dealers or by the local Department of Motor Vehicles. Therefore, strict anonymity cannot be enforced. However, it is possible to provide \emph{pseudonymity} by creating identifiers uncorrelated with real entities\cite{petit2014pseudonym}. The definition of a method to distribute identities and keys is beyond the scope of this paper.

\subsection{The Consensus Algorithm}
\label{sec:consensus}

\noindent For the consensus algorithm design, we evaluate established ones~\cite{nakamoto2008bitcoin, wood2014ethereum, kogias2016enhancing, king2012ppcoin, iost2017iost, nem2018nem, androulaki2018hyperledger} and customized them given the requirements of the automotive context.  
Consensus algorithms can be systematized into election-based or voting-based.

\emph{Election-based} algorithms rely on a resource that is difficult or expensive to obtain, like computational power or invested money. A leader, called the block harvester or the block miner, is then elected considering this resource. The harvester performs the validation of the block, digitally signs it, and broadcasts the resulting block to all the other participants, which only have to check the eligibility of the harvester to accept the new block. The main alternatives for this approach are \ac{PoW}, like in Bitcoin~\cite{nakamoto2008bitcoin}, and \ac{PoS} and its variants, like in Peercoin~\cite{king2012ppcoin}. \ac{PoW} can not be executed on embedded systems such as \ac{IVI} systems due to the computational requirements~\cite{o2014bitcoin}. On the other hand, \ac{PoS} is robust only if the resource that is at stake does not vary too quickly~\cite{daian2019snow}, which is not easily achievable in a network composed of vehicles where participants can join and leave frequently.

\emph{Voting-based} algorithms are variations of the \acf{PBFT}~\cite{castro1999practical} method. All the participants take part in the voting process and decide whether the candidate block proposed by the harvester is acceptable or not. 
Because of this feature, voting-based algorithms perform better in small networks, where they typically have higher throughput than election-based algorithms. Using this approach, all the participants execute the validation phase at least once per block. However, this overhead is undesirable because the framework would be much more resource-intensive. Moreover, the amount of exchanged messages required to achieve consensus is proportional to the square of the total number of participants, introducing significant overhead.

In blockchains, the election can be deterministic (e.g., ByzCoin\cite{jovanovic2016byzcoin}) or probabilistic (e.g., NEM\cite{nem2018nem}). 
A deterministic approach uses a synchronized view-number to perform the election. Instead, a probabilistic approach relies on a time-dependent condition evaluated locally by each participant to determine if it is eligible to harvest the next block. This means that, in this case, explicit synchronization among the participants is not required. However, several participants may try to harvest the next block simultaneously, causing the blockchain to fork, which does not occur in deterministic approaches. Eventually, a fork policy solves the issue, but the participants have already wasted their limited resources in discarded operations. 

\mypar{Proposed Consensus Algorithm}
In \systemname, we propose a consensus algorithm that combines election- and voting-based approaches adapting their functioning to the domain under analysis and finding a trade-off between resource efficiency and performance.
Instead of involving all the participants in the voting process, only a subset of them, which we defined as \emph{supporters}, perform the \ac{PBFT} algorithm on the new block. 
The harvester produces a new block; then, it broadcasts the new block to the rest of the participants, which stores the block and waits for its approval by the supporters.  
The supporters vote on whether to accept or reject the proposed block following the \ac{PBFT} method and broadcast their decisions, through approval signatures, to the non-supporting participants.
Once it receives enough approval signatures, each participant adds the corresponding block with the received approval signatures to its local copy of the blockchain. 
Unlike voting-based methods, the proposed consensus algorithm requires a robust method to determine when and how to elect the harvester and the supporters. For this reason, we introduce an election step inspired by election-based approaches.
Since probabilistic approaches may waste resources due to the simultaneous block harvesting and the consequential forking of the blockchain, we adopt a deterministic approach based on the elapsed time since the creation of the previous block. To avoid an easy to predict view-number, we derive its value from both the hash of the previous block -- on which there is already consensus -- and the elapsed time -- which can be kept synchronized with various solutions, e.g., \ac{GPS}. 
The harvester and the supporters are elected based on the view-number and their reputation. In a lottery-like fashion, each candidate receives several lottery tickets proportional to its reputation. Then, winning tickets (one for each supporter and harvester to be elected) are extracted based on the view-number. 
More formally, for each election, the view-number is concatenated with an increasing integer, and the resulting value is hashed using SHA-256 (i.e., winning tickets). The hash function is used as a uniform pseudo-random number generator. The algorithm then assigns to each node a partition of the image of the hash function used, proportionally to their reputation (i.e., number of lottery tickets per candidate). 
Finally, the algorithm elects the participant that owns the partition in which the hash is included. 
Since the output of the SHA-256 function is comparable to a uniform distribution, the probability of electing a participant is proportional to its reputation score. 
Regarding the number of elected participants, following the \ac{PBFT} method, we select a total of $n=3f+2$ candidates either as harvester or supporters, where $f$ represents the fault tolerance parameter, which corresponds to the minimum number of approval signatures required to add a new block. By doing so, we reduce the number of average validation phase executions per block to $n$.

\myparagraph{Proposed Algorithm Evaluation.} 
We demonstrate the robustness of the proposed consensus algorithm by analyzing the probability of a successful malicious block insertion in the case of a Sybil attack~\cite{douceur2002sybil}. To achieve the malicious goal, an attacker should be elected as harvester and control a number of supporters sufficient to win the voting procedure. Considering $3f+1$\footnote{$f$ is the fault tolerance parameter of \acf{PBFT}} supporters, a malicious block insertion is successful if the harvester node is malicious and the attackers own over 33\% of the supporter nodes. To perform the analysis, we model the problem by assigning a probability distribution to each involved variable. 
We define as $H$ the probability that an attacker has been elected as a block harvester, which is a Bernoulli random variable with success probability $p$, where $p$ is the ratio of the attackers' reputation with respect to the total reputation.
We define as $M$ the number of nodes controlled by attackers elected as supporters and will participate in the voting process. Since $n=3f+2$ is the total number of elected nodes between harvester and supporters and $f$ corresponds to the minimum number of approval signatures required to add a new block, we model $M$ as a binomial random variable with $3f+1$ trials, each having a success probability $p$, which is the same as in $H$. 
Equation~\ref{eq:attack-success-pr} shows how the formulas for the probability of success of attackers (i.e., malicious block insertion) are derived.

\small
\begin{align}
&H \sim Bernoulli(p) \nonumber\\
&M \sim Binomial(3f+1, p) \nonumber\\
&\begin{aligned}
Pr(&\text{``A malicious block is inserted''}) =\\
=& Pr(H = 1 \land M \geq f + 1) =\\
=& Pr(H = 1)(1-Pr(M \leq f)) =\\
=& p\cdot\left(1- \displaystyle\sum_{k=0}^{f}\binom{3f+1}{k}p^k(1-p)^{3f+1-k}\right) 
\label{eq:attack-success-pr}
\end{aligned} 
\end{align}
\normalsize

\begin{figure}
  \centering
    \includegraphics[width=0.4\textwidth]{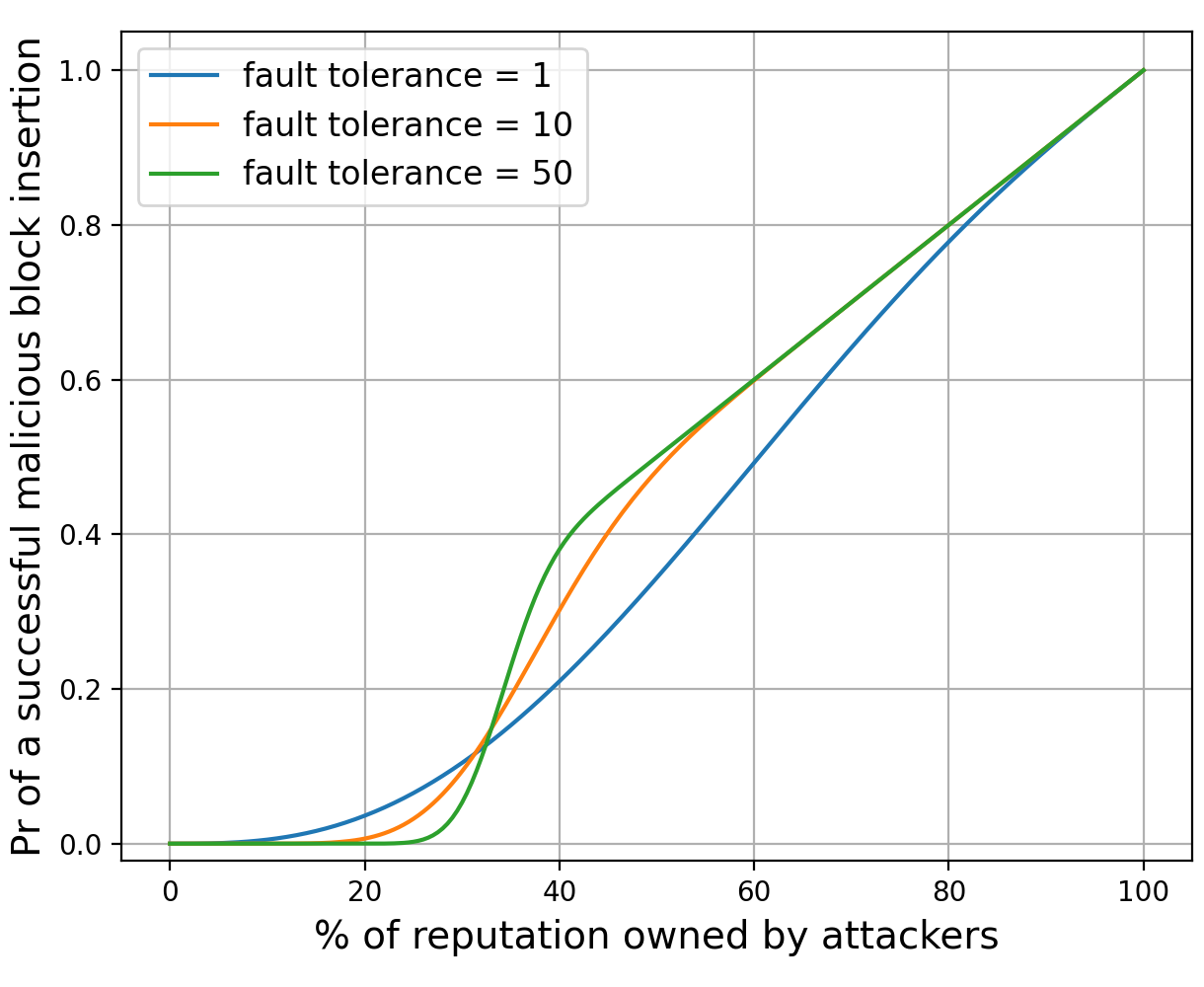}
  \caption{Probability of a successful malicious block insertion in relation with the amount of reputation owned by attackers.}
\label{fig:consensusattacks}
\vspace{-10 pt}
\end{figure}

To evaluate the robustness of the proposed approach, we, therefore, study the probability of a successful malicious block insertion by varying $p$ (i.e., as the amount of reputation owned by attackers grows) and $f$, as shown in Figure~\ref{fig:consensusattacks}. It is evident how by keeping $p$ in the interval $[0; 33\%]$ (i.e., the intersection between the $f$ curves), an increase in the fault tolerance parameter $f$ maintains the success rate of attackers low, if not close to zero.

\subsection{The Validation Algorithm}
\label{sec:validation}

\noindent In light of the strict computational requirements of \ac{IVI} systems,  we designed three heuristics that achieve the same result -- i.e., the validation of the block in the ledger performed by elected participants -- of algorithms implemented in contexts where the computation power is not limited (e.g., Bitcoin\cite{o2014bitcoin}).
In particular, in this phase, we validate a block by removing invalid transactions and updating participants' reputations following a \textit{reputation update strategy}.  

\subsubsection{Illegal transaction filtering} The first heuristic filters out illegal transactions and apply a penalty (i.e., lower participants' reputations) to their owners.
First, it filters out transactions that have the same target and sender IDs because the sender could otherwise exploit the validation algorithm to increase its reputation score by itself.
Then, it removes transactions that are not \enquote{acknowledged} by participants. In other words, for each transaction, there must be another transaction with sender and receiver IDs swapped, i.e., from the detected vehicle to the original sender. 
By doing so, we prevent a single attacker (i.e., not organized in a group) from masking itself as another vehicle.  

\subsubsection{Transaction validation} \label{ssec:trans-val}
The second heuristic validates each transaction individually. It considers the transactions contained in the previous $N$ blocks and checks if the information contained (i.e., position and timing) is compliant with the laws of physics. By assuming a linear motion, it is possible to find an upper bound to the travel distance of a vehicle for a given interval of time $\Delta t$ and a maximum velocity $v_{max}$. Therefore, we reject a transaction depending on participants' reputations and whether the travel distance exceeds these bounds. Also, we apply a penalty to transactions' owners. 
By comparing the identifiers contained in old and new transactions, we define four bounds, depending on whether the participant under analysis is sender or detected node in the transactions: 
the \emph{sender-sender bound},
the \emph{sender-detected bound},
the \emph{detected-sender bound},  
and \emph{detected-detected bound}. 
For the sender-sender bound, we compute the euclidean distance between the sender positions in different transactions since the values of its positions are directly indicated in the transaction.  
This is not true for all other bounds, where the node under analysis is a detected one. In these cases, the position of the detected node must be approximated based on the position of its sender, the geometry of the problem, and the maximum range of detection $range_{max}$. For example, the detected-sender bound is:
$$distance(sender_{old}, sender_{new}) \leq v_{max} \Delta t + range_{max}$$  
Then, we assign a score for each new transaction under analysis: 
when the travel distance is not coherent with the bounds, we lower the score assigned to the transaction of a value equal to the reputation of the sender of the older transaction. Otherwise, the score is increased by the same amount.
Finally, we normalize the score with respect to the number of bound checks made and compare it with a threshold. The new transaction is rejected if the normalized score is lower than the threshold. The evaluation of the transaction is not executed if the number of matches is equal to zero.

\subsubsection{Malicious groups penalization} \label{ssec:malicious-penalty}
The third heuristic penalizes the misbehavior of malicious groups, i.e., groups of participants that support each other to boost their reputations while faking localization by forging transactions. 
This heuristic is necessary since this behavior would pass unnoticed through the previous heuristics. 
We adopt a strategy to penalize misbehaving participants over time by analyzing each participant independently and detecting conflicts. There is a conflict between two participants whenever they claim to be in two close positions, but they do not detect each other (i.e., there is no such transaction within the block).
For each conflicting pair of participants, the one with the lowest reputation score receives a penalty. When a participant loses more than a defined amount of conflicts, all of its transactions are blacklisted. 
This strategy is based on the assumption that misbehaving participants have, on average, less reputation than legitimate ones. 
Finally, it is essential to highlight that this heuristic does not claim to provide an exact representation of the traffic stream but only approximates it by analyzing the transactions exchanged between participants. As previously explained, each block contains a limited number of transactions, and the remaining are dropped. Consequently, this heuristic may have a high number of false positives. 
However, experiments have shown that it does not deteriorate the overall performance since it is unlikely that a correctly behaving participant generates many conflicts systematically. 

\myparagraph{Reputation update strategy.}   
We update the reputation of the ledger participants based on the outcome of the previous heuristics and their activity in the blockchain. 
To do so, we first define as active a participant with at least one transaction in the block. 
The reputation of active participants is updated by summing to its original value the difference between their total reward and total penalty.
The total penalty is computed by multiplying a penalty base value with the number of rejected transactions. 
Instead, the total reward is computed by multiplying a reward base value with the number of accepted transactions and a scaling factor that considers the number of frequent neighbors (i.e., participants that often appear together in the transactions). The scaling factor mitigates the capability of groups of malicious participants to boost each others' reputation: the more a group of participants interacts only with each other, the lower the scaling factor and, hence, the reward is. The scaling factor is computed as: 
$$S_f =1-w\frac{|new_i\cap old_i|}{|old_i|}$$ 
where $w$ is the neighbor variance weight, $new_i$ is the set of participants detected by participant $i$ in the new block, and $old_i$ is the participants detected in the previous $M$ blocks. A low neighbor variance is more common in malicious groups, as they are only a few and require to interact frequently with each other to gain support and carry out the attack.
Regarding the reward and penalty base value, we select a fixed value over a dynamic one (i.e., a value that changes depending on the node history) since it performed better in our experiments. Dynamic value often zeroes the reputation of legit participants, making them unable to contribute to the validation of transactions. We also defined a maximum value for the reputation score so that participants that never go offline, such as \acp{RSU}, cannot gain infinite reputation and, hence, monopolize the execution of the consensus algorithm.
Inactive participants, instead, periodically lose their reputation to give more weight to recent transactions over older ones.
This guarantees that malicious participants have a lower reputation score than legit ones and that it is more difficult for them to increase it.

\section{Experimental Results}
\label{sec:experiments}

\noindent We evaluate  \systemname in a  simulated  (but realistic)  environment.
First, we analyze the performance and the robustness of the proposed framework through parameter studies. Second, we test the resistance of the framework against attacks derived from the threat model.

\subsection{Simulation Setup}

\noindent The experiments conducted on \systemname are carried out through simulations designed to be as realistic as possible and provide reliable results. 
For this purpose, we developed a modular simulator based on the Veins vehicular network simulation framework~\cite{veins}. This choice allowed us to use the OMNeT++\cite{omnetpp} simulation library to simulate the network model and SUMO\cite{sumo}, a microscopic traffic simulator.
Also, we adopt the \ac{MoST} scenario~\cite{codeca2017most} which is based on the Principality of Monaco and simulates 46,000 trips in 10 hours. This setup guarantees a realistic road layout, vehicle interactions, and traffic dynamics.
Our experiments consider 1,500 vehicles interacting for one hour to keep the computational and storage requirements manageable.  

\begin{table}
    \renewcommand{\arraystretch}{1.3}
    \centering
    \caption{Values of the parameters used in the experiments obtained through the performance and robustness analysis.}
    \label{tab:configuration}
    \begin{tabular}{|c|c|c|} \hline
         \bfseries Component & \bfseries Parameter & \bfseries Value  \\ \hline
         \multirow{3}{*}{Scenario} & No. of vehicles& 1,500\\ 
          & No. of \acp{RSU}& 40\\
          & Max speed & 130 km/h \\ \hline
         \multirow{2}{*}{Blockchain} & Blocksize & 50,000\\ 
          & Blocktime & 60 s\\ \hline
         \multirow{2}{*}{Transaction validation} & No. of previous blocks& 3\\ 
          & Threshold & 0.1\\ \hline
         \multirow{2}{*}{Misbehavior penalization} & Variance weight & 0.67\\ 
          & Tolerable lost conflicts & 2\\ \hline
         Consensus algorithm & Fault tolerance & 6\\ \hline
         \multirow{4}{*}{Reward system} & Max reputation& 4,096\\ 
          & Initial reputation & 64\\ 
          & Base reward & 256\\ 
          & Base penalty & 512\\ \hline
    \end{tabular}
    \vspace{-10pt}
\end{table}

\subsection{Experiment 1: Performance and Robustness Analysis}
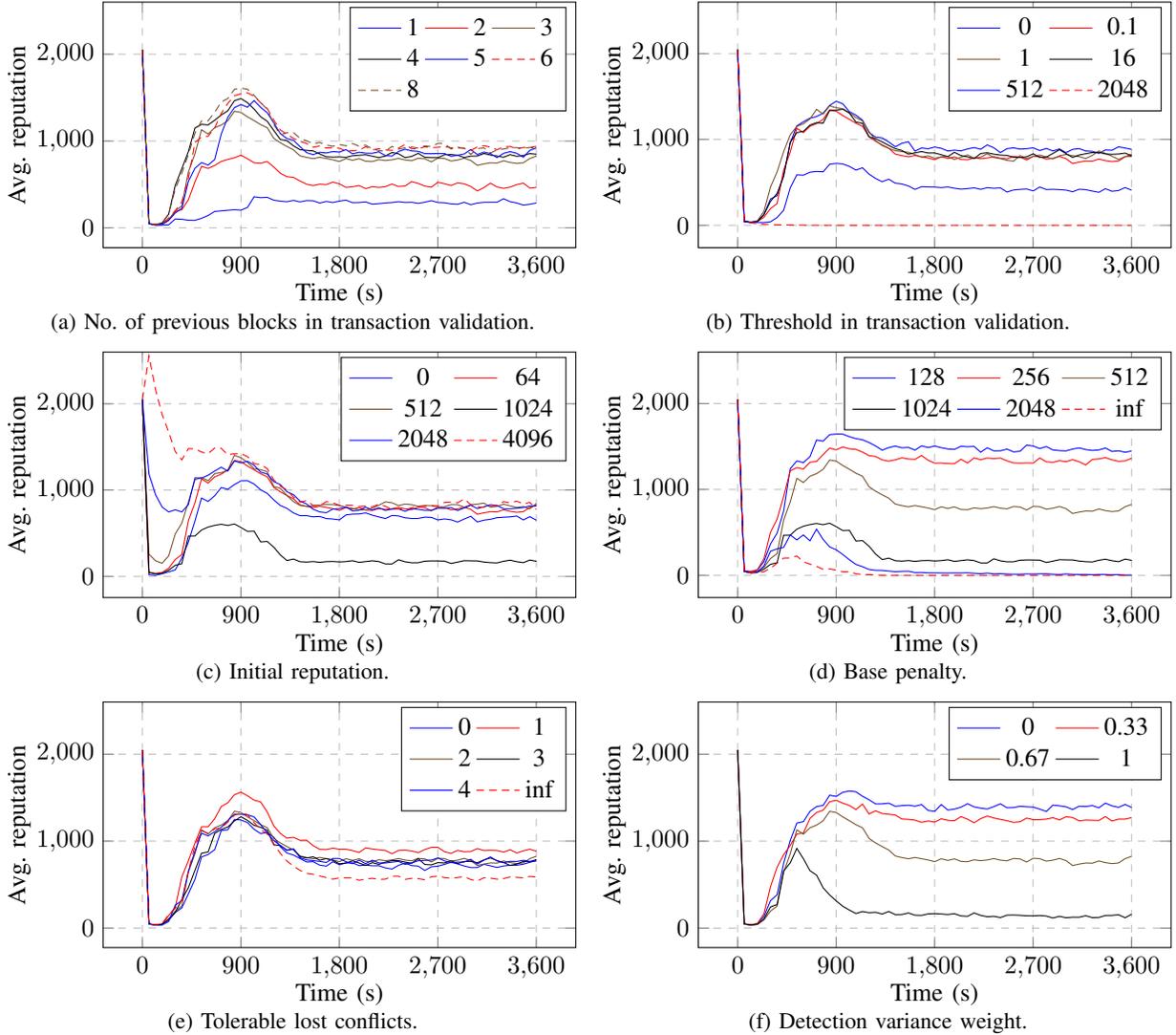
\begin{figure*}
    \centering
    \begin{subfigure}[b]{.45\textwidth}
    \begin{tikzpicture}
    \begin{myaxis}[3]
        \myparamstudyplot{experiments/prev-blocks/1.dat}{1}
        \myparamstudyplot{experiments/prev-blocks/2.dat}{2}
        \myparamstudyplot{experiments/prev-blocks/3.dat}{3}
        \myparamstudyplot{experiments/prev-blocks/4.dat}{4}
        \myparamstudyplot{experiments/prev-blocks/5.dat}{5}
        \myparamstudyplot{experiments/prev-blocks/6.dat}{6}
        \myparamstudyplot{experiments/prev-blocks/8.dat}{8}
    \end{myaxis}
    \end{tikzpicture}
    \vspace{-7pt}
    \caption{No. of previous blocks in transaction validation.}
    \label{fig:study-n}
    \end{subfigure}
    \begin{subfigure}[b]{.45\textwidth}
    \begin{tikzpicture}
    \begin{myaxis}
        \myparamstudyplot{experiments/threshold/0.0.dat}{0}
        \myparamstudyplot{experiments/threshold/0.1.dat}{0.1}
        \myparamstudyplot{experiments/threshold/1.dat}{1}
        \myparamstudyplot{experiments/threshold/16.dat}{16}
        \myparamstudyplot{experiments/threshold/512.dat}{512}
        \myparamstudyplot{experiments/threshold/2048.dat}{2048}
    \end{myaxis}
    \end{tikzpicture}
    \vspace{-7pt}
    \caption{Threshold in transaction validation.}
    \label{fig:study-threshold}
    \end{subfigure}
    \par\medskip
    \begin{subfigure}[b]{.45\textwidth}
    \begin{tikzpicture}
    \begin{myaxis}
        \myparamstudyplot{experiments/initial-reputation/0.dat}{0}
        \myparamstudyplot{experiments/initial-reputation/64.dat}{64}
        \myparamstudyplot{experiments/initial-reputation/512.dat}{512}
        \myparamstudyplot{experiments/initial-reputation/1024.dat}{1024}
        \myparamstudyplot{experiments/initial-reputation/2048.dat}{2048}
        \myparamstudyplot{experiments/initial-reputation/4096.dat}{4096}
    \end{myaxis}
    \end{tikzpicture}
     \vspace{-7pt}
    \caption{Initial reputation.}
    \label{fig:study-initial-reward}
    \end{subfigure}
    \begin{subfigure}[b]{.45\textwidth}
    \begin{tikzpicture}
    \begin{myaxis}[3]
        \myparamstudyplot{experiments/penalty/128.dat}{128}
        \myparamstudyplot{experiments/penalty/256.dat}{256}
        \myparamstudyplot{experiments/penalty/512.dat}{512}
        \myparamstudyplot{experiments/penalty/1024.dat}{1024}
        \myparamstudyplot{experiments/penalty/2048.dat}{2048}
        \myparamstudyplot{experiments/penalty/inf.dat}{inf}
    \end{myaxis}
    \end{tikzpicture}
    \vspace{-7pt}
    \caption{Base penalty.}
    \label{fig:study-penalty}
    \end{subfigure}
    \par\medskip
        \begin{subfigure}[b]{.45\textwidth}
    \begin{tikzpicture}
    \begin{myaxis}
        \myparamstudyplot{experiments/max-conflicts/0.dat}{0}
        \myparamstudyplot{experiments/max-conflicts/1.dat}{1}
        \myparamstudyplot{experiments/max-conflicts/2.dat}{2}
        \myparamstudyplot{experiments/max-conflicts/3.dat}{3}
        \myparamstudyplot{experiments/max-conflicts/4.dat}{4}
        \myparamstudyplot{experiments/max-conflicts/inf.dat}{inf}
    \end{myaxis}
    \end{tikzpicture}
     \vspace{-7pt}
    \caption{Tolerable lost conflicts.}
    \label{fig:study-conflicts}
    \end{subfigure}
    \begin{subfigure}[b]{.45\textwidth}
    \begin{tikzpicture}
    \begin{myaxis}
        \myparamstudyplot{experiments/variance-weight/0.0.dat}{0}
        \myparamstudyplot{experiments/variance-weight/0.33.dat}{0.33}
        \myparamstudyplot{experiments/variance-weight/0.67.dat}{0.67}
        \myparamstudyplot{experiments/variance-weight/1.0.dat}{1}
    \end{myaxis}
    \end{tikzpicture}
    \vspace{-7pt}
    \caption{Detection variance weight.}
    \label{fig:study-weight}
    \end{subfigure}
    \caption{Experiment 1: Analysis of the average reputation of active participants, varying one parameter at a time.}
    \label{fig:study}
    \vspace{-10 pt}
\end{figure*}

\noindent The framework needs to have a stable and robust configuration that allows it to function correctly. For this reason, in the first experiment, we analyze the impact of the \systemname parameters to empirically estimate their optimal values and their sensitivity in a context where no attack occurs. For each of the parameters under analysis, we analyze the average reputation of participants (i.e., the stake of our \ac{PoS}) as a metric of the goodness of our system. In particular, we run different simulations varying the value of the parameters\footnote{In the following experiments, we highlight only the most relevant values}.  In Table~\ref{tab:configuration}, we show the final values obtained for each parameter that we then use for subsequent experiments. 

\myparagraph{Number of previous blocks for transaction validation. }
First, we study the number of preceding blocks (and their related transactions) used to validate new transactions and, therefore, recognize the attacks. The higher is the number of blocks used, the more the transaction validation algorithm complexity and the computation time increase. On the other hand, the lower is the number of blocks used, the higher is the chance of rejecting legitimate transactions (due to the lack of a robust transaction history), lowering the reputation of legitimate nodes. Figure~\ref{fig:study-n} shows the average reputation of active participants under different values of this parameter. When the framework reaches the steady-state, all but two configurations produce similar outcomes, suggesting that using more than three previous blocks does not improve the transaction validation enough to justify the increased complexity. 

\myparagraph{Threshold in transaction validation. }
The threshold used in the transaction validation (see Section \ref{ssec:trans-val}) is necessary to define the minimum required support to accept a new transaction. A threshold too high would compromise the system's functioning since it would lead to a high number of legitimate rejected transactions (i.e., the score assigned to new transactions would rarely satisfy the threshold) and, therefore, an increase in the penalization of legitimate participants. Figure~\ref{fig:study-threshold} shows a comparison of the average reputation as the value of the threshold varies. 
Low threshold values (i.e., 0, 0.1, 1, 16) produce similar outcomes (i.e., the average reputation shows an almost identical trend and a low variance). 
However, as the thresholds increase in value, the average reputation drops, as expected. In particular, the highest tested threshold (i.e., 2048) completely prevents the correct functioning of the framework, forcing the average reputation to drop to zero abruptly. Considering a scenario where no attacks occur, a low value of the threshold guarantees the highest performance. Consequently, we choose a threshold of 0.1. The goodness of the threshold is also confirmed by results obtained in experiment 2.

\myparagraph{Initial reputation.}
The initial reputation given to participants when they join the system for the first time should not significantly affect the proposed framework's long-term performance. A high initial reputation would lead to a conveniently fast convergence to a steady-state. However, this configuration could be exploited by attackers that, with a high initial amount of reputation, would have higher chances of succeeding with their attacks. Instead, the convergence to a steady-state is slower with a low initial reputation, but attackers would not have a multiplier factor to exploit since there would be no difference between new participants and untrusted ones. The results of this analysis are presented in Figure~\ref{fig:study-initial-reward}, which show that there is no significant difference in the steady-state by selecting different values of initial reputation. Only the first ten minutes of the simulation are significantly affected. The configuration with no initial reputation, however, exhibits worse behavior than the others. This is caused by the fact that the participants with zero reputations cannot support others' transactions, causing more transactions to get rejected. For this reason, 
we opt for a low but non-zero value of the initial reputation (i.e., 64).

\myparagraph{Reward/Penalty base values.}
The most relevant parameters in the reward system are the $reward base value$ and $penalty base value$. The $reward base value$ should be significantly smaller than the $penalty base value$ to positively reward a consistently correct behavior and heavily penalize misbehaviors, thus reducing the chances of attack success. To study the effects of these two parameters, we fixed the $reward base value$ -- to 256 -- and let the $penalty base value$ vary. 
Figure~\ref{fig:study-penalty} displays the comparison of the outcomes of the simulations. As expected, higher penalties result in a lower average reputation. 

However, when the penalty is too high, the average reputation decreases significantly, increasing the chances of a transaction being rejected and leading to the worst-case scenario in which the reputation of participants becomes zero. In this case, attackers can exploit the low average reputation of legitimate participants to carry out a successful attack. Therefore, excessively high penalties are not advisable. On the other side, as we tested on attack scenarios, a base penalty not high enough does not penalize the attackers for their misbehavior, enabling them to send false information before being ignored. We choose a value of 512 for the base penalty, which has proven to be a good trade-off value.

\myparagraph{Maximum conflicts for the malicious group penalization.}
The malicious group penalization simulation must be carefully analyzed to prevent the framework from collapsing into undesirable states. The parameter that drives the simulation is the number of tolerable lost conflicts, which determines when to blacklist a participant based on its behavior. If too many lost conflicts are allowed, malicious behavior is more likely to pass unnoticed. It is essential although to consider that spontaneous conflicts may arise naturally. However, our simulations suggest that the number of spontaneous conflicts is, in general, low. 

The average reputation of active participants under different values of this parameter is shown in Figure~\ref{fig:study-conflicts}, where it is possible to notice that the least restrictive condition produces the lowest average reputation.  This is because if transactions are blacklisted, the evaluation of other transactions is avoided due to the lack of support. Therefore, overall, less penalty is distributed because rejected transactions are not used to detect conflicts. For this reason, we choose two as the maximum number of conflicts before being blacklisted.  

\myparagraph{Reputation update strategy variance weight.}
The weight given to the neighbors' variance directly affects the variation of overall reputation (see Section~\ref{ssec:malicious-penalty}).
A low neighbor variance is typical of cooperative malicious since they communicate with each other only to increase their reputation.   
However, also legitimate participants can interact with the same vehicles or \acp{RSU} more than once, especially in case of traffic congestion. 
Figure~\ref{fig:study-weight} confronts the average reputations obtained by changing the neighbor variance weight. When the framework reaches a steady-state, the average reputations under the different configurations are distinct and well-separated. This shows that even in non-attack scenarios, neighbors often repeat transactions between each other and that the weight needs to be tuned to obtain a good average reputation. Therefore, a trade-off between usability (i.e., low weight value) and security (i.e., high weight value) must be considered.
Since it remains essential to ensure that attackers do not feed each other reputation, we choose 0.67 as an in-between weight that keeps the average reputation high enough while not rewarding repeated neighbor detection.

\subsection{Experiment 2: Resistance to Attacks}

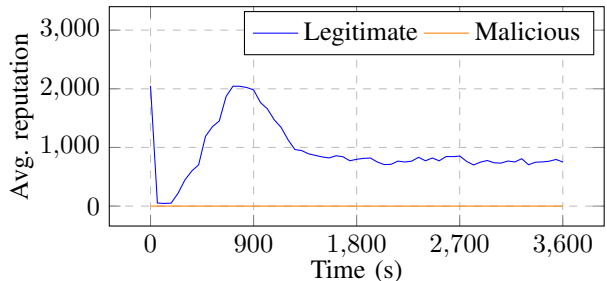
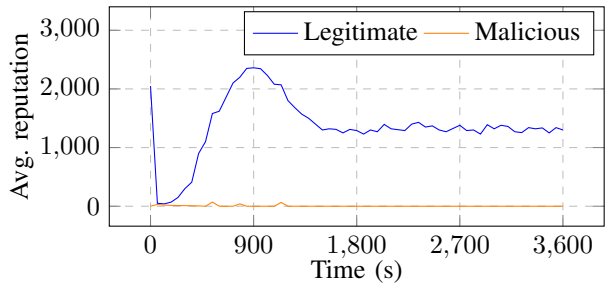
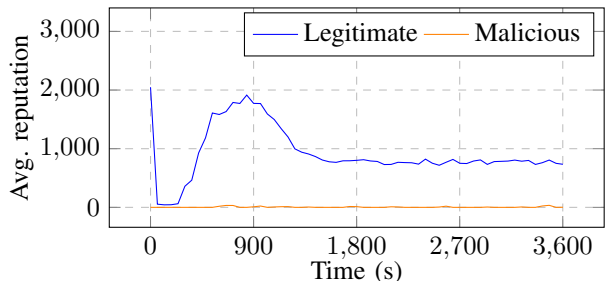
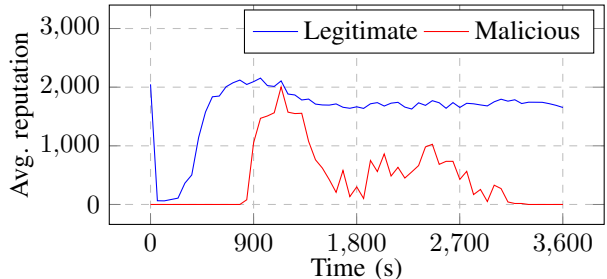
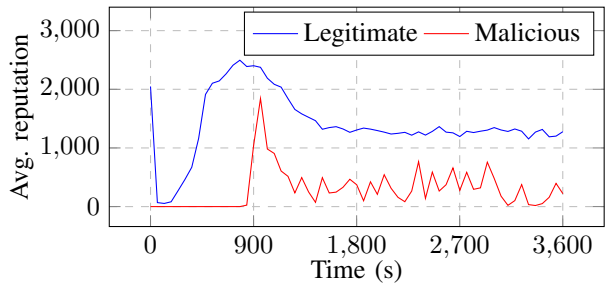
\begin{figure}

    \begin{subfigure}[b]{.45\textwidth}
    \begin{tikzpicture}
    \begin{myaxis1}
        \myattackplot{experiments/attacks/random-attack.dat}
    \end{myaxis1}
    \end{tikzpicture}
    \vspace{-7pt}
    \caption{Randomly generated transaction attacks.}
    \label{fig:attack-random}
    \end{subfigure}
    
    \vspace{7pt}
    
    \begin{subfigure}[b]{.45\textwidth}
    \begin{tikzpicture}
    \begin{myaxis1}
        \myattackplot{experiments/attacks/shift-attack.dat}
    \end{myaxis1}
    \end{tikzpicture}
    \vspace{-7pt}
    \caption{Offset position transaction attacks.}
    \label{fig:attack-shift}
    \end{subfigure}
    
    \vspace{7pt}
    
    \begin{subfigure}[b]{.45\textwidth}
    \begin{tikzpicture}
    \begin{myaxis1}
        \myattackplot{experiments/attacks/mimic-attack.dat}
    \end{myaxis1}
    \end{tikzpicture}
    \vspace{-7pt}
    \caption{Replayed transaction attacks.}
    \label{fig:attack-mimic}
    \end{subfigure}
    
    \vspace{7pt}
    
    \begin{subfigure}[b]{.45\textwidth}
    \begin{tikzpicture}
    \begin{myaxis1}
        \mycoordinatedattackplot{experiments/attacks/self-sustaining-group-5.dat}
    \end{myaxis1}
    \end{tikzpicture}
    \vspace{-7pt}
    \caption{Self-sustaining group of 5 members.}
    \label{fig:attack-group-1}
    \end{subfigure}
    
    \vspace{7pt}
    
    \begin{subfigure}[b]{.45\textwidth}
    \begin{tikzpicture}
    \begin{myaxis1}
        \mycoordinatedattackplot{experiments/attacks/self-sustaining-group-20.dat}
    \end{myaxis1}
    \end{tikzpicture}
    \vspace{-7pt}
    \caption{Self-sustaining group of 20 members.}
    \label{fig:attack-group-2}
    \end{subfigure}

    \caption{Experiment 2: Analysis of the average reputation of active legitimate and  malicious participants in different attack scenarios.}
    \label{fig:attack}
\end{figure}

\begin{figure*}
\centering
\includegraphics[width=.65\linewidth]{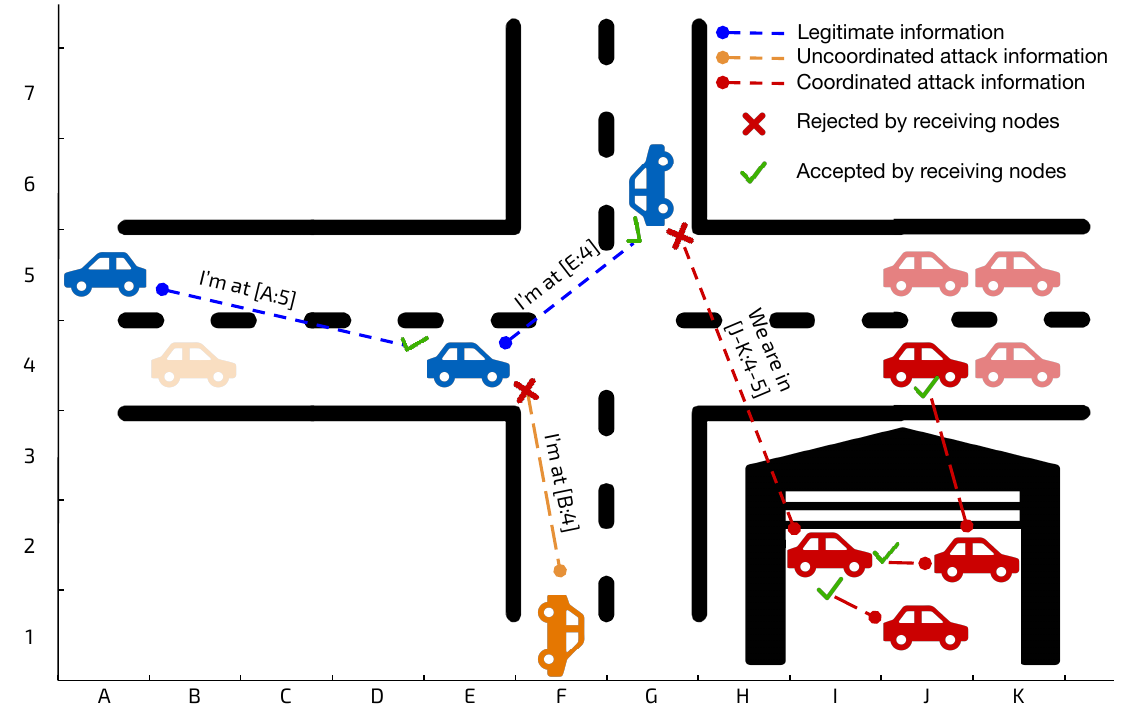}
    \caption{Experiment 2: Graphical representation of attack scenarios. In {\color{blue} blue} legitimate participants, in {\color{orange} orange} an uncoordinated attacker, and in {\color{red} red} coordinated attackers. Transparent vehicles represent the fake positions claimed by attackers.}
    \label{fig:attack-scenarios}
    \vspace{-10 pt}
\end{figure*}

\noindent In this experiment, we analyze the four attack scenarios modeled from the threat model described in Section~\ref{sec:background}: three of them involve uncoordinated attackers that act individually, while the last attack involves a coordinated group of attackers. We analyze each attack independently since they do not affect each other. For each attack scenario, we analyze the value of the average reputation of legitimate and malicious participants as a metric of the goodness of our system. In Figure~\ref{fig:attack-scenarios}, we represent a graphical representation of the attack scenarios. 

\begin{table}[b]
    \renewcommand{\arraystretch}{1.3}
    \centering
    \caption{Experiment 2: Analysis of the resilience of \systemname in terms of the maximum percentage of attackers that can be managed by our framework. For each attack scenario, we show the \% of accepted malicious block transactions of the successful attack.}
    \label{tab:maxattackers}
    \begin{tabular}{|c|c|c|} \hline
         \bfseries Attack Scenario & \bfseries \% of attackers & \bfseries \% of blocksize \\\hline
         
         Random (1)& 30\% & 26,80\% \\ \hline
         Position Changing (2a,b)& 80\% & 80\% \\ \hline 
         Replay (3) & 45\% & 16,4\% \\ \hline
         \end{tabular}
\end{table}

\myparagraph{Uncoordinated attacks scenarios.} We considered three attack scenarios that require no coordination among the attackers and no particular interactions with legitimate participants. 

The first attack involves the generation of a significant amount of valid-looking transactions with random sender position and detected vehicle/\ac{RSU} identifier to exhaust the space in the block. 
In our experiments, attackers generate and broadcast three times the transactions of the average user. Figure~\ref{fig:attack-random} shows the result of the simulation of this attack scenario. The average reputation of attackers is consistently zero, while legitimate participants have, on average, enough reputation for supporting correct traffic information.
As visible in Table~\ref{tab:maxattackers}, as long as the number of attackers does not exceed 30\% of the total number of users, \systemname is resilient to this attack. This is coherent with the threat model, which considers a relatively small number of attackers, and with the limitations of similar solutions that apply \ac{PBFT}-based algorithms, which requires that for $m$ attackers, at least $3m+1$ total nodes exist to ensure stability.

The second attack involves altering the sender's position to spoof a real position or lower the quality of traffic information. 
Regarding altering the sender's position, attackers add a constant offset vector to their real positions. 
Regarding lowering the quality of traffic information, the offset is changed for each spoofed transaction so that the attacker appears to be moving unrealistically. 
Both attacks generate similar results, shown in Figure~\ref{fig:attack-shift}, which follow a similar pattern to the first attack. 
As shown in Table \ref{tab:maxattackers}, \systemname resists these attacks as long as the number of attackers is lower than 80\% of the total users.

The third uncoordinated attack consists in replaying the content of the transactions generated by legitimate participants. In this scenario, attackers wait for legit participants to broadcast their transactions to copy the content except for the sender's identifier. Then, the attackers sign and broadcast the copied transaction. By doing so, an attacker can impersonate a legitimate participant, spoofing its position.
Figure~\ref{fig:attack-mimic} shows the outcome of the attack. Similar to the previous scenarios, the attack fails, and the reputation of the attackers remains zero with small spikes during the initial transient. The copied transactions usually are not confirmed by the detected participant, as the attacker is at another position and, hence, filtered by the illegal transactions filtering process of the validation algorithm. As in the previous attacks, the minimum number of attackers required for the framework to not detect the attacks is significantly higher than the one considered in the threat model, as visible in Table \ref{tab:maxattackers}.

\myparagraph{Coordinated attack scenario.}
In this attack scenario, attackers operate as a group. One of the attackers is on-road while the others are in a different position and behave properly until the attack begins.
At this point, the malicious group generates fake transactions (between each other), pretending to be close to the one on-road. 
The on-road attacker eventually stops interacting with the rest of the group, which now has enough reputation to sustain itself autonomously. Then, the self-sustaining malicious group generates fake traffic congestion by changing the advertised position. This attack affects the usability of \systemname even if a few attackers can succeed consistently.
Figure~\ref{fig:attack-group-1} shows the outcome of the attack when the self-sustaining malicious group is composed of 5 members. After an initial peak, the average reputation of the attackers starts dropping and eventually reaches zero. Consequently, the attack partially succeeds only for a limited period. The average reputation of the attackers is kept lower than the one of legitimate participants thanks to both the misbehavior penalization heuristic and the reputation update heuristic. Figure~\ref{fig:attack-group-2} shows the same attack with 20 attackers. Although the attackers do not reach zero reputation during the simulated time, their reputation remains very low and easily distinguishable from the legitimate ones. The number of accepted malicious transactions is less than 50\% of the malicious ones in the block, implying that the attack remains contained although not completely blocked.

\section{Conclusions} 
\label{sec:conclusion}

\noindent This paper presents \systemname, a blockchain-based decentralized framework that can be used as an alternative to widespread centralized solutions to collect real-time vehicular information. We demonstrated its feasibility in the context of vehicle positioning and traffic information management. Consequently, we defined a threat model tailored to the automotive domain. We validated \systemname by implementing a network simulator for realistic traffic generation. The results show that the framework can function stably in realistic scenarios while also maintaining high-quality traffic information even in the presence of malicious entities.

One of the limitations of this work is related to the limited availability of real traffic dynamics. Therefore, we resorted to use the SUMO traffic framework and the MoST scenario to produce a realistic simulation. However, MoST contains only one sequence of simulated events. To avoid overfitting, we evaluated our framework in different time windows of the MoST scenario. 
Another limitation is the computational complexity of the consensus mechanism, which has a quadratic dependency on the number of transactions stored in a block. Therefore, the proposed framework may not scale well with big blocks. The quadratic dependency is determined by the fact that every transaction in the new block has to be compared with all the transactions contained in older blocks or the new blocks themselves. Although it is possible to parallelize the critical loops of the validation algorithm, the computational complexity is still quadratic on the size of the block and cannot be reduced without redesigning the algorithms. 
An alternative approach and future work is to rely on a statistical approach for the consensus mechanism, where a node has only to check a smaller and fixed portion of the blocks. This approach could effectively reduce the computational complexity to linear, but it would carry the challenge of probabilistic guarantees that need to be carefully proved and enforced.

Overall, this research lays the foundations for further investigation toward decentralized blockchain-based vehicular systems, since to the best of our knowledge, there is no proposal for a decentralized traffic monitoring framework that can be used in a real environment.

\IEEEpeerreviewmaketitle

\section*{Acknowledgments}

\noindent This project has been partially supported by BVTech SpA under project grant UCSA.

\bibliographystyle{IEEEtran}
\bibliography{Bibliography}

\begin{IEEEbiography}[{\includegraphics[width=1in,height=1.25in,clip,keepaspectratio]{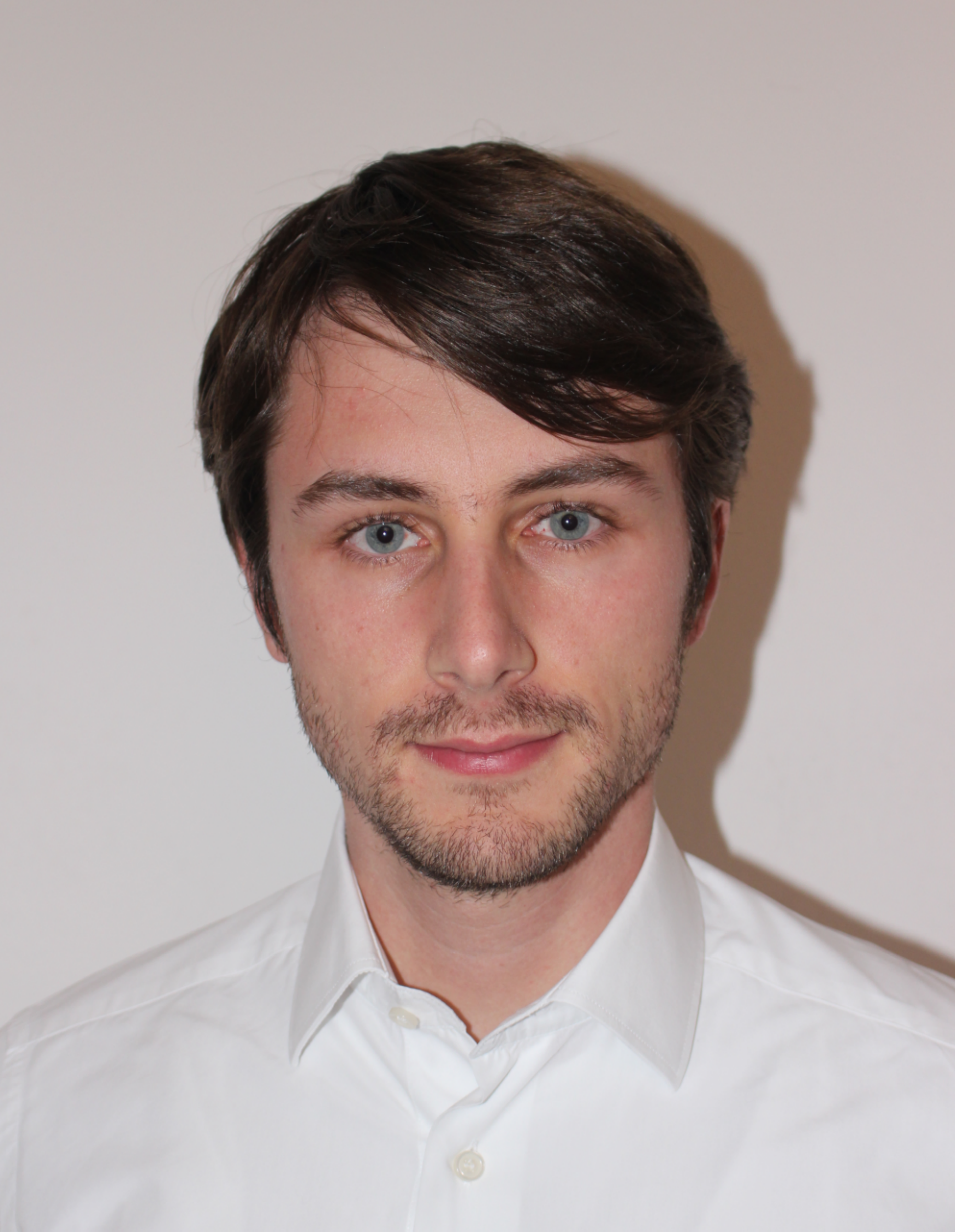}}]{Davide Maffiola}
Davide Maffiola is a graduated student from Politecnico di Milano. There, he earned both his BSc and his MSc with honors in Computer Science and Engineering in 2018 and 2020 respectively. His main interests are in Computer Network Security, Embedded Systems, and Cyber-Physical Systems Security. He is currently working in the area of Risk Advisory.
\end{IEEEbiography}
\begin{IEEEbiography}[{\includegraphics[width=1in,height=1.25in,clip,keepaspectratio]{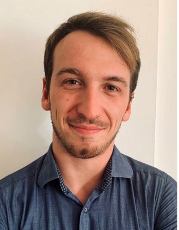}}]{Stefano Longari}
Stefano Longari received a Ph.D. in Information Technology from Politecnico di Milano, where he currently works as a Postdoctoral Researcher in NECST laboratory as part of the System Security group inside the Dipartimento di Elettronica, Informazione e Bioingegneria. His main research focus is automotive security and the use of novel technologies such as machine learning intrusion detection to secure the development of smart mobility and smart city ecosystems.
\end{IEEEbiography}
\begin{IEEEbiography}
[{\includegraphics[width=1in,height=1.25in,clip,keepaspectratio]{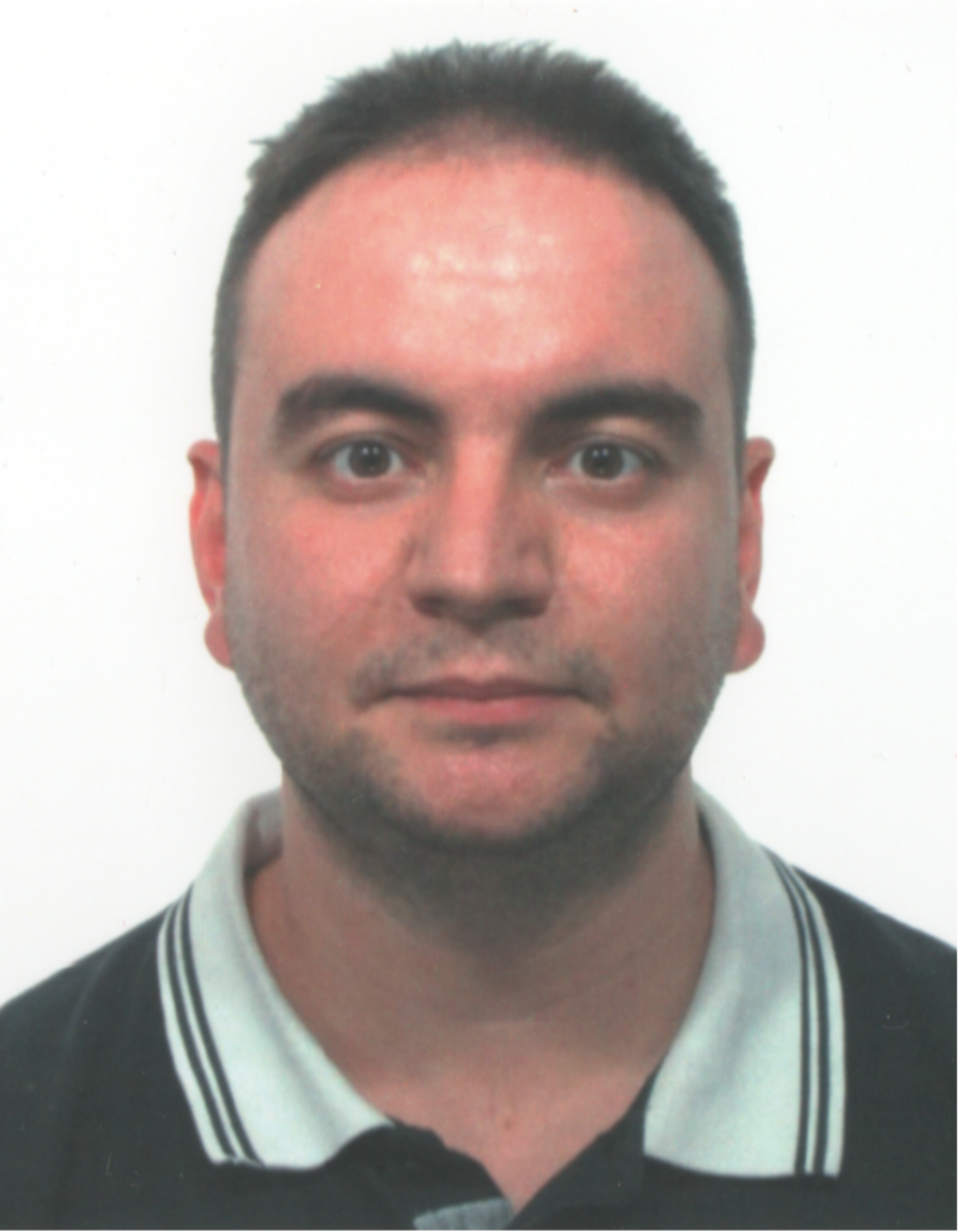}}]{Michele Carminati}
Michele Carminati received his Ph.D. degree cum laude in Information Technology from Politecnico di Milano in Italy, where he is currently a Contract Professor and a Postdoctoral Researcher working at NECST laboratory as part of the System Security group inside the Dipartimento di Elettronica, Informazione e Bioingegneria. His research revolves around the application of machine learning methods in various cybersecurity-related fields, ranging from cyber-physical and automotive systems to binary and malware analysis, going through anomaly and intrusion detection. He is actively involved in research projects funded by the European Union, and he is also co-founder of Banksealer, a Fintech spinoff of Politecnico di Milano.
\end{IEEEbiography}
\begin{IEEEbiography}
[{\includegraphics[width=1in,height=1.25in,clip,keepaspectratio]{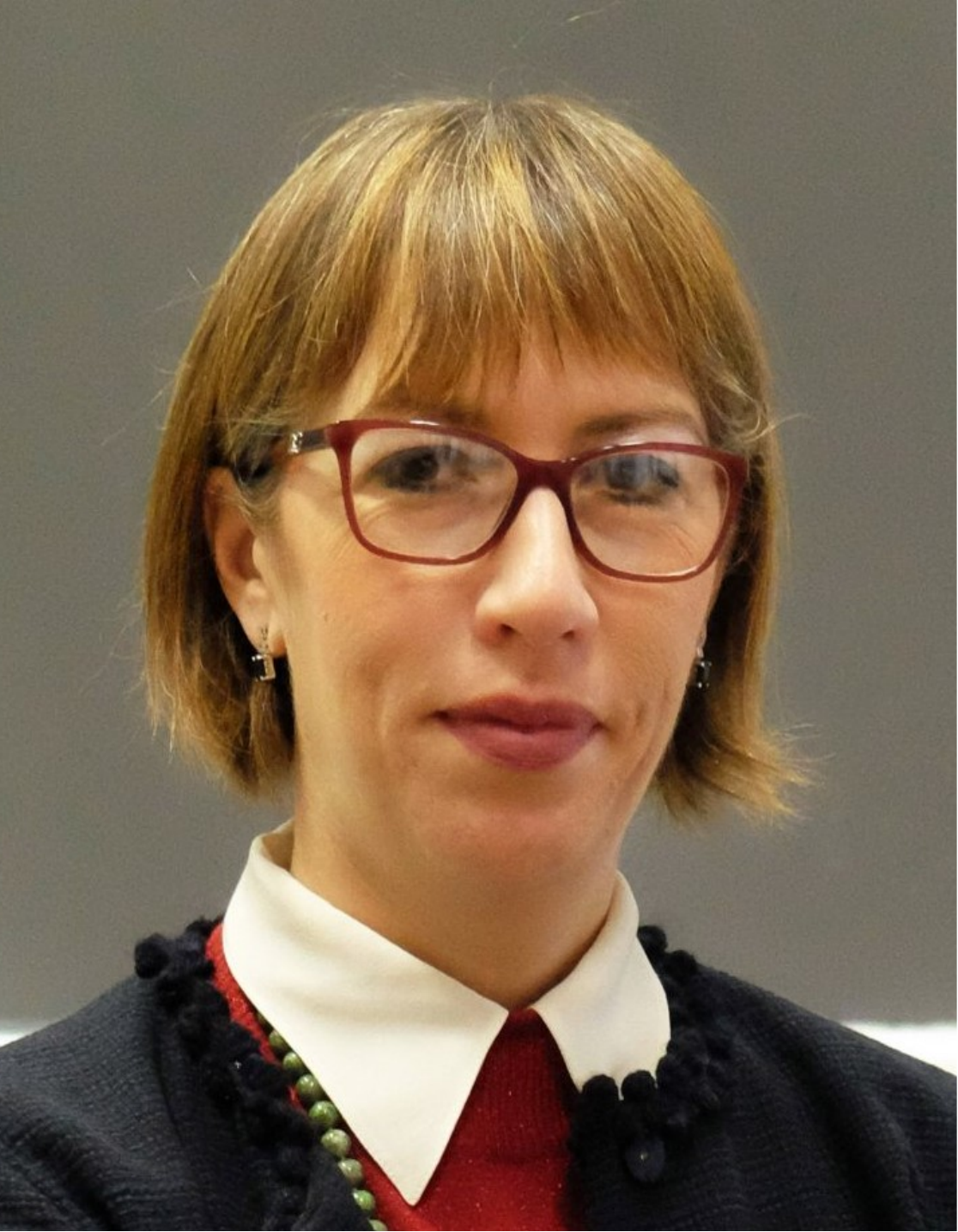}}]{Mara Tanelli}
Mara Tanelli (M'05, SM'12) is currently a Full Professor of Automatic Control at the Politecnico di Milano, where she earner her Ph.D. in Information Engineering with honors in 2007. Her main research interests are in Automotive Control, Smart Mobility and Industrial Analytics. She is co-author of more than 170 peer-reviewed publications in these research areas. Prof. Tanelli is a member of the Conference Editorial Board of the IEEE Control Systems Society (CSS). She is AE for the IEEE  Transactions on Control Systems Technology and for the IEEE Transactions on Human-Machine Systems. Since 2021 she is Chair of the Technical Committee (TC) on Automotive Controls of the IEEE CSS and vice-chair for publications of the corresponding IFAC TC.
\end{IEEEbiography}
\begin{IEEEbiography}
[{\includegraphics[width=1in,height=1.25in,clip,keepaspectratio]{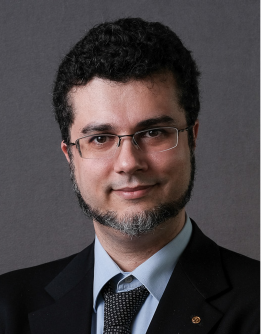}}]{Stefano Zanero}
Stefano Zanero (M’03,  SM’10) received a Ph.D. degree in Computer Engineering from
Politecnico di Milano, where he is currently an Associate Professor with
the Dipartimento di Elettronica, Informazione e Bioingegneria. His
research focuses on cybersecurity, in particular cyber-physical systems,
malware analysis, and fraud and anomaly detection through the use of
machine learning. Prof. Zanero is a Distinguished Contributor of the IEEE Computer Society, where he currently serves on the Board of Governors.
\end{IEEEbiography}

\end{document}